\documentclass[a4paper,fleqn]{cas-dc}

\usepackage[authoryear,longnamesfirst]{natbib}

\usepackage{algorithmicx}
\usepackage{algorithm}
\usepackage{algpseudocode}

\usepackage{mathrsfs}

\def\tsc#1{\csdef{#1}{\textsc{\lowercase{#1}}\xspace}}
\tsc{WGM}
\tsc{QE}

\begin{document}
\let\WriteBookmarks\relax
\def\floatpagepagefraction{1}
\def\textpagefraction{.001}

\shorttitle{}    

\shortauthors{}  

\title [mode = title]{A Physics-Embedded Dual-Learning Imaging Framework for Electrical Impedance Tomography}



%

\author[1,2]{Xuanxuan Yang}
\author[1]{Yangming Zhang}
\author[3]{Haofeng Chen}
\author[2]{Gang Ma}
\author[1]{Xiaojie Wang}

\affiliation[1]{
    organization={Hefei Institutes of Physical Science, Chinese Academy of Sciences},
    addressline={350 Shushanhu Road},
    city={Hefei},
    postcode={230031},
    state={Anhui},
    country={China}
}

\affiliation[2]{
    organization={Department of Precision Instruments and Precision Machinery, University of Science and Technology of China},
    addressline={96 Jinzhai Road},
    city={Hefei},
    postcode={230026},
    state={Anhui},
    country={China}
}

\affiliation[3]{
    organization={Department of Cybernetics, Faculty of Electrical Engineering, Czech Technical University in Prague},
    addressline={Karlovo náměstí 13},
    city={Prague},
    postcode={121 35},
    state={},
    country={Czech Republic}
}

\fntext[preprint]{This is the author's accepted manuscript of the article:
X. Yang, Y. Zhang, H. Chen, G. Ma, and X. Wang,
``A Physics-Embedded Dual-Learning Imaging Framework for Electrical Impedance Tomography,''
\textit{Neural Networks}, p.~108464, 2025. 
DOI: \href{https://doi.org/10.1016/j.neunet.2025.108464}{https://doi.org/10.1016/j.neunet.2025.108464}. 
© 2025 Elsevier. This manuscript is made available under the arXiv.org perpetual, non-exclusive license.
The final published version is available at the above DOI link.}

















\begin{abstract}
Electrical Impedance Tomography (EIT) is a promising noninvasive imaging technique that reconstructs the spatial conductivity distribution from boundary voltage measurements. However, it poses a highly nonlinear and ill-posed inverse problem. Traditional regularization-based methods are sensitive to noise and often produce significant artifacts. Physics-Embedded learning frameworks, particularly Physics-Informed Neural Networks (PINNs), have shown success in solving such inverse problems under ideal conditions with abundant internal data. Yet in practical EIT applications, only sparse and noisy boundary measurements are available. Moreover, changing boundary excitations require the simultaneous training of multiple forward networks and one inverse network, which significantly increases computational complexity and hampers convergence. To overcome these limitations, we propose a Physics-Embedded Dual-Learning Imaging Framework for EIT. The dual-learning strategy is composed of a supervised CNN-based forward network, which learns to predict a discrete internal potential distribution under fixed Neumann-to-Dirichlet boundary conditions, and an unsupervised PINN-based inverse network, which reconstructs the conductivity by enforcing the governing PDE through discrete numerical differentiation of the predicted potentials. This decoupled architecture removes the need for smooth conductivity assumptions, reduces the number of forward networks required from $K$ to 1, and improves reconstruction robustness and efficiency under realistic measurement constraints.(https://github.com/XuanxuanYang/CNN-PINNframework.git)
\end{abstract}


\maketitle

\section{Introduction}
Electrical Impedance Tomography (EIT)\cite{cheney1999electrical} is a form of tomographic imaging that calculates the spatial distribution of internal conductivity by injecting currents and measuring boundary voltages. EIT has broad applications in fields such as medical imaging \cite{bodenstein2009principles}, multiphase flow detection\cite{xxy2022realtime}, and tactile sensing\cite{thurner2024dynamic}. Solving EIT involves a nonlinear partial differential equation (PDE) derived from Maxwell’s equations and presents a computationally challenging, ill-posed inverse problem. Physics-Informed Neural Networks (PINNs) have recently emerged as a machine learning approach for solving PDEs\cite{raissi2019physics,cuomo2022scientific}. By embedding additional physical constraints into the loss function, PINNs approximate PDE solutions by minimizing the physics-informed loss during training. This approach, known as Learning with Physics Loss\cite{guo2023physics}, offers strong interpretability and has demonstrated significant advantages for solving inverse problems\cite{mishra2022estimates,baldan2023physics}. Consequently, researchers have started exploring the use of PINNs to enable tomographic imaging, including applications to EIT\cite{bar2021strong,pokkunuru2023improved,ruan2024magnetic}.

However, existing research on using PINNs for EIT has primarily focused on the semi-inverse EIT problem rather than the full-inverse EIT problem. Following the terminology introduced by Bar et al.\cite{bar2021strong} and Pokkunuru et al.\cite{pokkunuru2023improved}, we define these two settings as: 

\begin{itemize}
\item \textbf{Semi-inverse EIT:} Given full access to the internal potential field $u(x,y)$, the goal is to reconstruct the conductivity distribution $\sigma(x,y)$ within the ROI $\mathscr{H}$, i.e., $u(x,y)\mapsto\sigma(x,y)$. This formulation is mathematically convenient but unrealistic, since internal measurements are generally unavailable in practice.
\item \textbf{Full-inverse EIT:} Only boundary voltage measurements $\Delta V$ are available, and the task is to reconstruct $\sigma(x,y)$ from these sparse data, i.e., $\Delta V\mapsto\sigma(x,y)$. This scenario is considerably more challenging due to its ill-posedness and the limited amount of available information. 
\end{itemize}

L. Bar et al.\cite{bar2021strong} were the first to apply PINNs for semi-inverse EIT, demonstrating promising results under the smooth conductivity assumption. Subsequently, A. Pokkunuru et al.\cite{pokkunuru2023improved} introduced data-driven energy-based priors\cite{lecun2006tutorial} to accelerate PINN convergence and enhance imaging accuracy. Solving semi-inverse EIT involves two networks: a forward network and an inverse network. The forward PINN network $f_{u}^{semi}\left( x,y \right) $ predicts the potential $u(x,y)$ at any point $(x,y)\in \mathscr{H} \subset \mathbb{R} ^2$, given the known conductivity distribution within the target domain $\mathscr{H} $. The inverse PINN network $f_{\sigma}^{\mathrm{semi}}(x,y)$, on the other hand, estimates the conductivity $\sigma (x,y)$ across $\mathscr{H} $, given the internal potential field $u(x,y)$ at all points. Automatic differentiation(AD) is then applied to $f_{u}^{\mathrm{semi}}(x,y)$ to compute the gradient, which serves as a constraint in the PDE loss for the inverse network.This approach requires $f_{u}^{\mathrm{semi}}(x,y)$ to be explicitly differentiable, allowing the use of tools like TensorFlow’s $tf.gradients$ function. This smoothness assumption\cite{evans2010partial} underpins the Gaussian-smoothing used in these studies, restricting conductivity distributions to simple shapes, such as circles or ellipses.

But in practice, we often face the full-inverse EIT problem. Although more challenging, efforts have been made by L. Bar et al. and A. Pokkunuru et al. in this area. A. Pokkunuru et al. acknowledged that their method could only reconstruct simple shapes (circles or ellipses), and failed when faced with more complex distributions. They also attempted to replicate L. Bar et al.’s approach but were unsuccessful, commenting, “To the best of our knowledge, only Bar $\&$ Sochen (2021) claim to train the EIT inverse problem using PINNs by jointly training $u$-Net and $\sigma$-Net, but unfortunately, their implementation is not open-source and we were unable to reproduce their results based on the details in the paper.” This suggests that true full-inverse EIT remains unachieved due to the need for simultaneous training of multiple forward networks and a single inverse network, which presents significant convergence challenges.

Given these challenges, the limitations of existing works are
as follows:
\begin{enumerate}
\item Current methods\cite{bar2021strong,pokkunuru2023improved} are restricted to the idealized semi-inverse EIT scenario and are not yet capable of addressing full-inverse EIT under realistic conditions.
\item To ensure explicit differentiability of the forward network $f_{u}^{semi}\left( x,y \right)$, existing methods rely on the assumption that conductivity is smooth and continuous\cite{evans2010partial}.
\item In full-inverse EIT, the boundary conditions vary with each change in excitation source position. When the source position changes $K$ times, joint training of $K$ forward networks and a single inverse network is required. This simultaneous training imposes dependency across all $K$+1 networks in terms of parameters, loss functions, and gradient updates, which strengthens the coupling between forward and inverse networks\cite{bar2021strong,pokkunuru2023improved,guo2023physics}. Consequently, this approach results in increased computational costs, memory usage, time, and convergence difficulties.
\end{enumerate}

To address the limitations discussed above, we propose a hybrid imaging framework that combines CNN and PINN, decoupling the forward and inverse problems to enable practical application of PINNs in full-inverse EIT. Recently, CNNs have become indispensable in image processing due to their ability to capture spatial features and invariant patterns, demonstrating exceptional performance in tasks such as image segmentation\cite{alom2018recurrent}, regression\cite{kolotouros2019convolutional}, and object detection\cite{gidaris2015object}. Observing that potential distributions exhibit spatial characteristics similar to images, we leverage CNNs to capture the relationship between boundary voltage measurements and internal potential distributions. Thus, we construct a data-driven, supervised forward network using CNNs to output a discrete potential distribution $\mathscr{U}_d$ under fixed Neumann-to-Dirichlet (NtD) boundary conditions for any given boundary voltage measurement $\Delta \mathrm{V}$. Subsequently, we design a model-driven, unsupervised inverse network using PINNs, which obtains the necessary PDE constraints via discrete numerical differentiation of the forward network’s output $\mathscr{U}_d$. This enables the reconstruction of the conductivity distribution $\Sigma$ from the discrete potential field $\mathscr{U}_d$, effectively bridging the data-driven and physics-driven components of the framework.

Furthermore, unlike traditional PINN-based EIT, which requires idealized continuous boundary current sources and smooth conductivity assumptions, our approach introduces a CNN-based forward predictor to handle realistic point-electrode excitations and non-smooth conductivity distributions. The CNN component serves as a data-driven prior that enables robust potential estimation under sparse and noisy boundary measurements, while the subsequent PINN stage enforces governing physics and eliminates the pure black-box nature. This hybrid mechanism reduces the reliance on over-simplified mathematical assumptions and paves the way toward future practical hardware-ready EIT imaging systems. 

This study builds upon our preliminary work reported in a short letter(four pages) \cite{yang2025two}, where we first introduced the idea of integrating a CNN-based forward model with a PINN-based inverse model for EIT. That earlier version served as an initial proof-of-concept but was limited in scope, lacking detailed mathematical formulations, in-depth comparisons, and experimental validation.
In contrast, the present paper provides a comprehensive extension, including:
(i) rigorous mathematical definitions of both semi-inverse and full-inverse problems,
(ii) the introduction of discrete numerical differentiation to decouple forward and inverse networks and overcome the smoothness assumptions in previous PINN approaches,
(iii) the adoption of realistic electrode-based boundary conditions instead of continuous idealized ones,
(iv) systematic investigation of how different boundary conditions affect reconstruction accuracy,
(v) robustness analysis under varying levels of synthetic and experimental noise,
(vi) evaluation of extreme conductivity contrasts (very high/low values) to examine the stability of the proposed method in challenging regimes,
(vii) provision of algorithmic pseudocode to improve transparency and reproducibility of our framework, and
(viii) extensive validations through both simulations and real phantom experiments.

The contributions of this paper are as follows:

\begin{enumerate}
\item Proposes a physics-embedded two-stage framework that enables PINNs to perform full-inverse EIT reconstruction under sparse, realistic measurements by combining a supervised CNN-based forward model with an unsupervised physics-constrained inverse model.

\item Replaces automatic differentiation with discrete numerical differentiation, removing smoothness assumptions on conductivity and decoupling forward and inverse problems, thus reducing the number of forward networks from $K$ to 1.

\item Adapts the framework to practical 16-electrode point-excitation experiments, and validates performance through both challenging non-smoothed conductivity phantoms with sharp boundaries and real water-tank experiments. 
\end{enumerate}

\section{Mathematical Framework and Comparison of EIT Approaches }

\subsection{PDE-based formulation of EIT }
The PDE forms the foundation of PINNs, and we will now use it to provide a clearer description of EIT, which reconstructs the spatial conductivity $\sigma$ in target domain $\mathscr{H} \subset \mathbb{R}^2$ based on voltage difference measurements by electrodes at the boundary $\partial \mathscr{H}$. The distribution of the potential $u$ within the ROI is governed by an elliptic equation:
\begin{equation}
\label{deqn_ex1a}
-\nabla \cdot (\sigma \nabla u) = 0 \quad \text{in } \mathscr{H} \subset \mathbb{R}^2
\end{equation}

To initiate the EIT process, a current $\zeta$ is applied to the surface $\partial \mathscr{H}$ via electrodes at the boundary, and the resulting voltage $u|_{\partial \mathscr{H}} = \kappa$ on $\partial \mathscr{H}$ is measured. Under these conditions, the Neumann and Dirichlet boundary conditions are defined as:
\begin{align}
\sigma \left( \frac{\partial u}{\partial n} \right) =\zeta \quad \text{on } \partial \mathscr{H}
\\u|_{\partial \mathscr{H}}=\kappa \quad \text{on } \partial \mathscr{H}
\end{align}
where $n$ represents the outward unit normal vector on $\partial \mathscr{H}$. Together, these conditions define the Neumann-to-Dirichlet (NtD) operator:
\begin{equation}
\label{deqn_ex1a}
\Lambda _{\sigma}:\zeta \mapsto \kappa
\end{equation}
where $\Lambda _{\sigma}$ maps the applied current $\zeta$ to the measured boundary voltage $\kappa$, providing the mathematical basis for reconstructing conductivity $\sigma$.

EIT reconstructs the spatial conductivity distribution \(\sigma(x, y)\) in \(\mathscr{H}\) based on boundary voltage measurements \(\partial \mathscr{H}\). The mathematical foundation of EIT lies in solving the PDE described by Equation (1), governed by an elliptic operator with Neumann and Dirichlet boundary conditions. These conditions, defined by the Neumann-to-Dirichlet (NtD) operator, establish the relationship between the applied current \(\zeta\) and the resulting boundary voltage \(\kappa\).  

However, traditional PINN-based methods (TPM)\cite{bar2021strong,pokkunuru2023improved} and our proposed approach differ significantly in their approach to Neumann boundary conditions, conductivity assumptions, and handling of discrete data. Below, we mathematically formalize these differences to highlight the advantages of the proposed approach in addressing full-inverse EIT challenges.

\begin{figure}[htbp]
\centering
\includegraphics[scale=0.6]{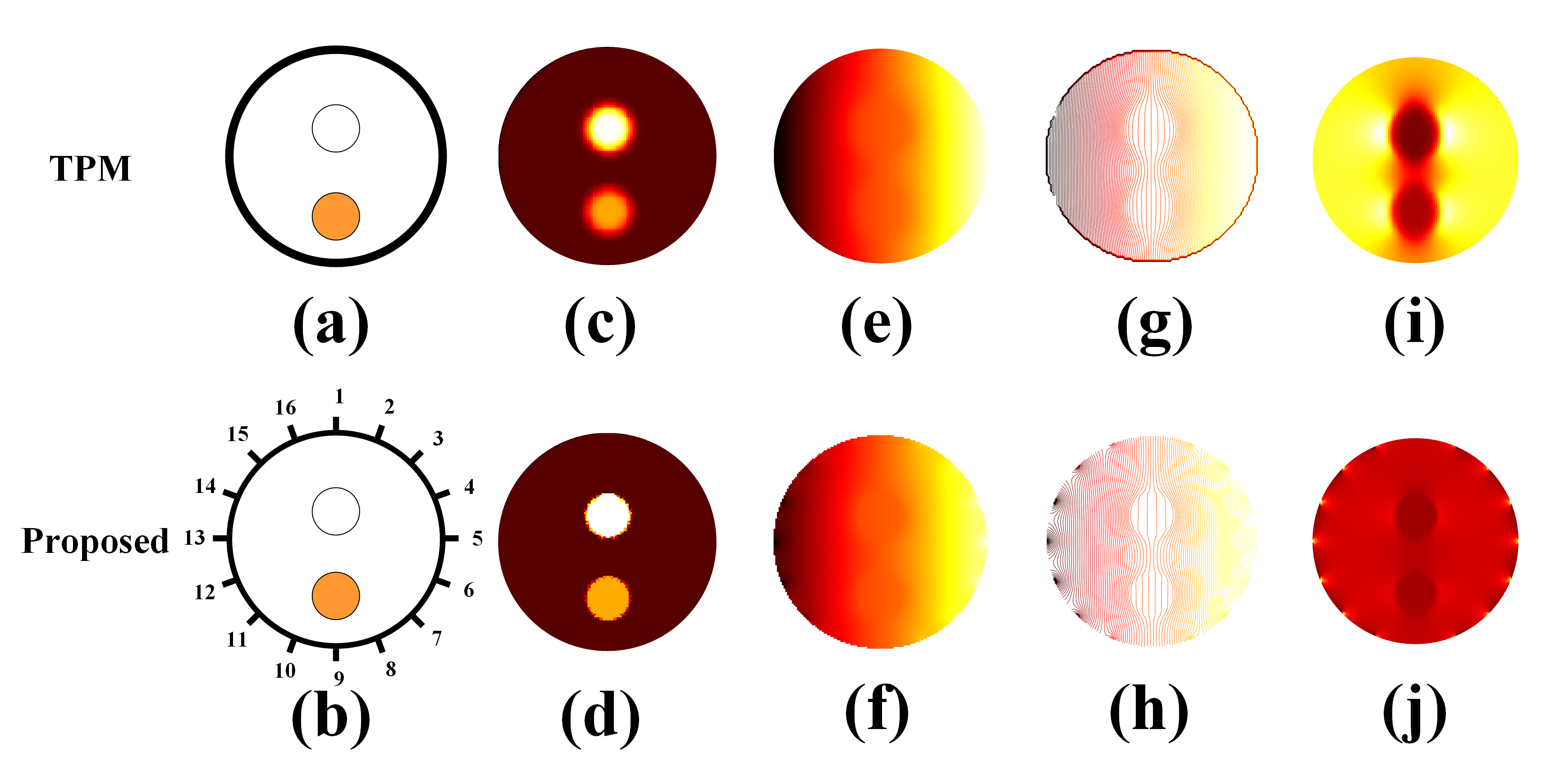}
\caption{(a) Continuous excitation on the boundary.  
(b) Excitation by electrodes on the boundary.  
(c) Conductivity distribution with Gaussian low-pass filtering.  
(d) Conductivity distribution without Gaussian low-pass filtering.  
(e, f) Electric potential field.  
(g, h) Contour of the electric potential field.  
(i, j) Partial derivative of the potential distribution with respect to the $x$-direction.}
\label{TPMvsOurs}
\end{figure}

\subsection{Neumann condition based on finite electrodes}
In TPMs, the Neumann boundary condition is set by applying continuous excitation along the entire boundary, as illustrated in Fig. \ref{TPMvsOurs}(a). This excitation is typically controlled by a trigonometric function\cite{siltanen2000implementation}, which generates a potential field in the internal region upon current injection, shown in Fig. \ref{TPMvsOurs}(c) with its corresponding contour plot in Fig. \ref{TPMvsOurs}(d).

While continuous boundary excitation can enhance imaging quality, it is an idealized setup. To evaluate the performance of the proposed approach under conditions closer to real-world scenarios, we apply excitation using 16 discrete electrodes along the boundary, as shown in Fig. \ref{TPMvsOurs}(b). The excitation function is defined as
\begin{align}
{\zeta =\frac{1}{\sqrt{2\pi}}\sin \left( \omega k+\varphi \right)}
\end{align}
where the frequency $\omega ={\frac{\pi}{8}}$, $k$ represents the electrode index, with $k\in \mathbb{Z} \text{ , }1\leqslant k\leqslant 16$, and the phase $\varphi =0$. The corresponding potential field and contour plot are shown in Fig. \ref{TPMvsOurs}(g) and (h), respectively.

\subsection{Handling non-smooth conductivity distributions}
In traditional TPMs, experiments are conducted under the assumption of smooth conductivity (Fig. 1(b)). This is because automatic differentiation requires the forward network $u(x,y)=f_u^{semi}(x,y)$ to be continuously differentiable in order to compute PDE losses in the inverse network. Consequently, the conductivity must be strictly positive and continuously differentiable, i.e., $\sigma(x,y)\in C^1(\mathscr{H})$ with $0<l_b\leq\sigma(x,y)$ \cite{evans2010partial}. However, such assumptions rarely hold in practice. Realistic conductivity distributions are often piecewise continuous, as shown in Fig. 1(f), where sharp transitions exist across object boundaries. To address this gap, we relax the smoothness assumption and allow $\sigma(x,y)$ to be piecewise continuous, thereby enabling the reconstruction of complex shapes such as triangles and squares with sharp edges, as well as validation using real experimental data.

\subsection{Discrete numerical differentiation for approximating continuous fields}
\begin{figure}[htbp]
\centering
\includegraphics[scale=0.8]{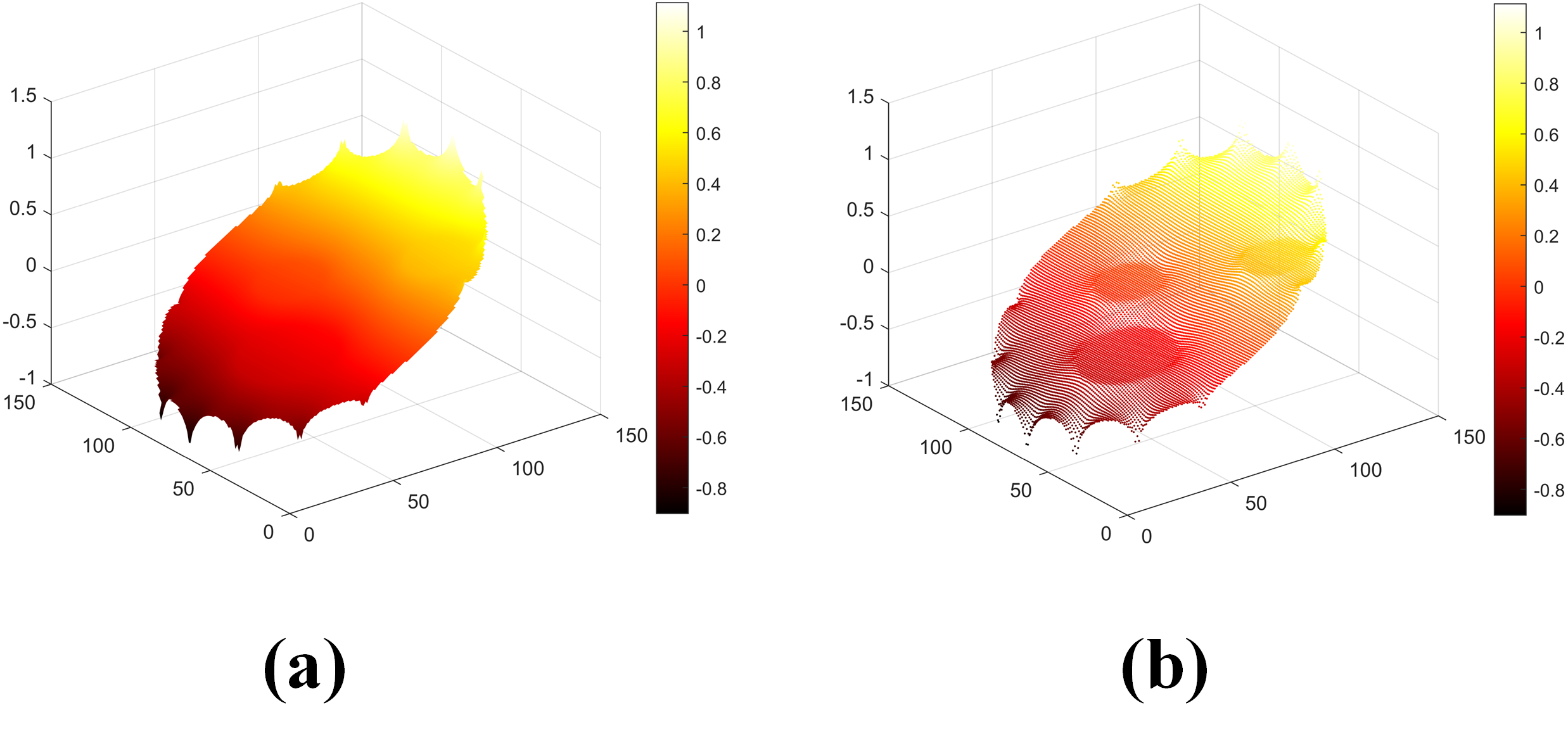}
\caption{(a)Continuous potential distribution. (b)Discrete potential distribution.}
\label{discrete vs continuous}
\end{figure}
To overcome the reliance on smoothness, we adopt discrete numerical differentiation on the potential field. Instead of requiring $u(x,y)$ to be differentiable everywhere, we define the forward solution on a discrete grid $\mathscr{D}\subset \mathbb{Z}^2$ and use a CNN-based forward network to predict the discrete potential distribution $\mathscr{U}_d={u_d(x,y)\mid(x,y)\in\mathscr{D}}$ on a $128\times128$ mesh. Gradients are then approximated using central finite differences, e.g.,
$\partial u_d/\partial x \approx \big(u_d(x+1,y)-u_d(x-1,y)\big)/(2h)$ and
$\partial u_d/\partial y \approx \big(u_d(x,y+1)-u_d(x,y-1)\big)/(2h)$,
where $h$ is the grid spacing.

As illustrated in Fig.~\ref{discrete vs continuous}, the output of the traditional TPM framework (a continuous potential field requiring global smoothness) is contrasted with that of our proposed approach (a discrete grid-based potential distribution). This discrete formulation not only avoids the need for global differentiability, but also allows CNNs to efficiently capture spatial dependencies in the potential field, thereby enabling the reconstruction of non-smooth conductivity distributions.

\begin{figure*}[htbp]
\centering
\includegraphics[scale=1]{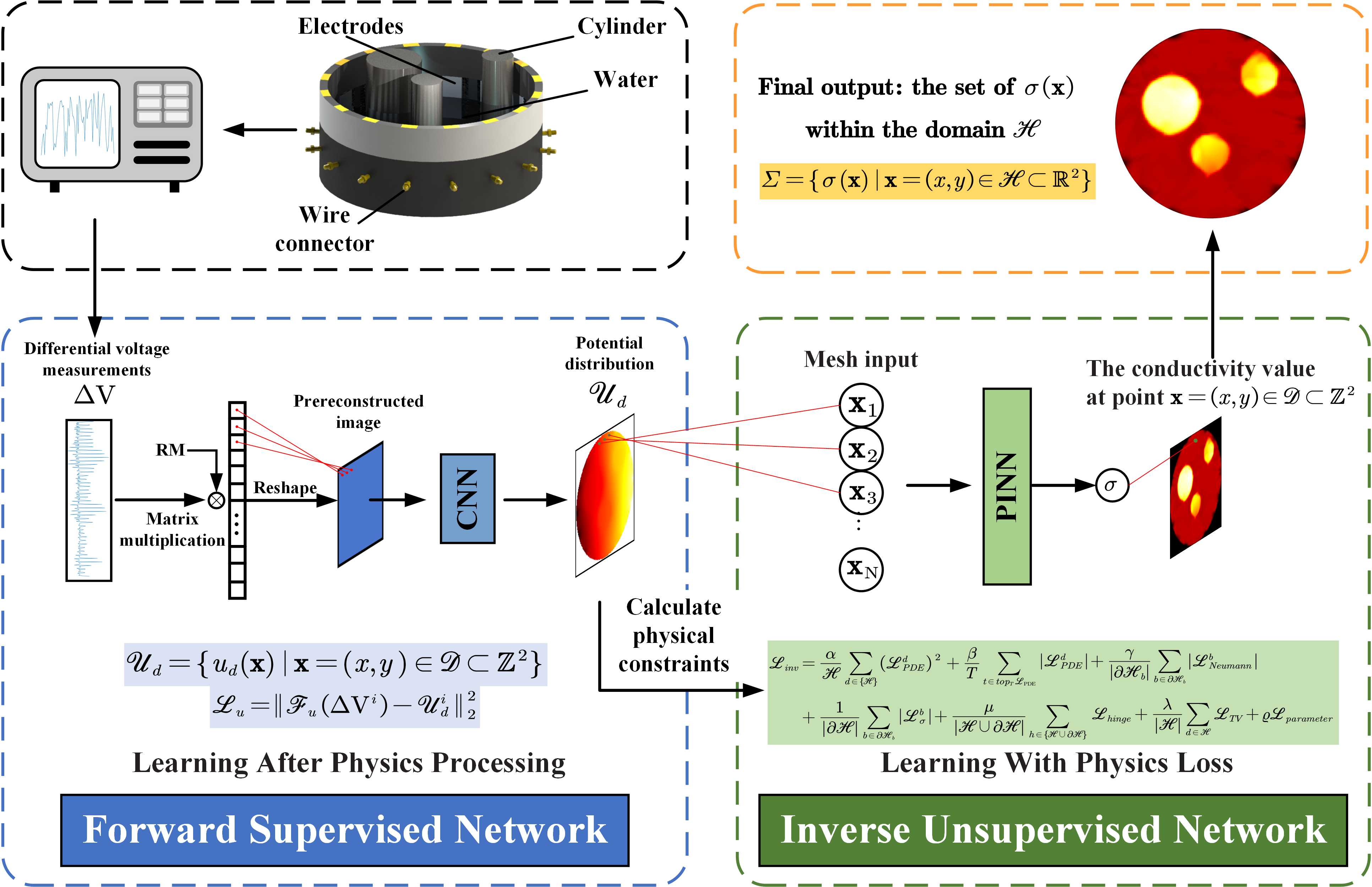}
\caption{The architecture of the proposed approach 
}
\label{CP-EIT}
\end{figure*}
\section{Method}

The overall framework of proposed approach is illustrated in Fig. \ref{CP-EIT}, comprising the EIT data acquisition module and the proposed approach module. The proposed approach consists of a forward supervised network $\mathscr{F}_u$ and an inverse unsupervised network $\mathscr{F}_\sigma$. First, the EIT data acquisition system collects the boundary voltage measurements $\Delta \mathrm{V}$, which are then fed into the forward supervised network $\mathscr{F} _u$ to generate a discrete potential distribution $\mathscr{U}_d$. Subsequently, discrete numerical differentiation is applied to guide the PDE loss term of the inverse unsupervised network $\mathscr{F} _\sigma$, enabling the reconstruction of the conductivity distribution $\varSigma $.

\subsection{Forward supervised network}
\begin{figure}[htbp]
\centering
\includegraphics[scale=0.7]{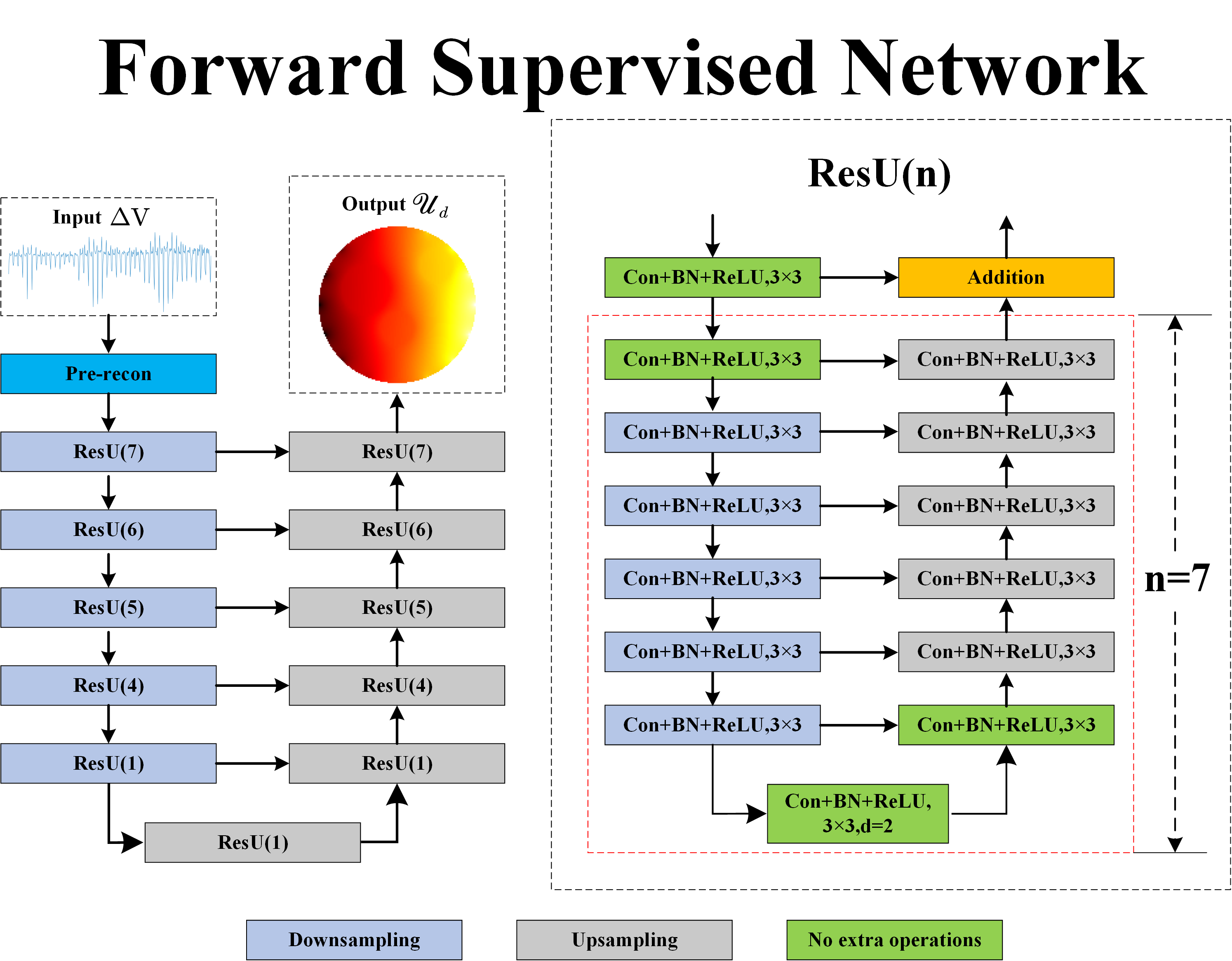}
\caption{The architecture of
forward supervised network.}
\label{F-NET}
\end{figure}
In this section, we provide a detailed introduction to the data-driven supervised forward network $\mathscr{F} _u$, as illustrated in Fig. \ref{F-NET}. The goal of $\mathscr{F}_u$ is to train a mapping from the boundary differential voltage measurements $\Delta \mathrm{V}$ to the discrete potential distribution $\mathscr{U}_d$ under fixed boundary conditions constrained by Equation (2). This training is performed using the dataset $\left\{ \Delta \mathrm{V}^i, \mathscr{U}_d^i \right\}_{i=1}^N$, and the problem can be formulated as:
\begin{align}
\hat{\theta}=\mathrm{arg}\min_{\theta} \frac{1}{N}\sum_{i=1}^N{\mathscr{L} _u\left( \mathscr{F} _u\left( \Delta \mathrm{V}^i \right) ,\mathscr{U} _{d}^{i} \right)}
\end{align}
where $\theta = \left\{ W, b \right\}$ represents the weights and biases of the forward network $\mathscr{F}_u$, and $\mathscr{L}_u$ is the loss function.
Since this is a regression problem, we adopt the mean squared error (MSE) loss, defined as:

\begin{align}
\mathscr{L} _u\left( \mathscr{F} _u\left( \Delta \mathrm{V}^i \right) ,\mathscr{U} _{d}^{i} \right) =\left\| \mathscr{F} _u\left( \Delta \mathrm{V}^i \right) -\mathscr{U} _{d}^{i} \right\| _{2}^{2}
\end{align}

\subsubsection{Physics-based processioning strategy}
When boundary currents are injected, the current experiences abrupt changes as it flows through objects, significantly influencing the potential distribution, as shown in Fig. \ref{TPMvsOurs}(f). This effect is particularly evident in the spatial gradient of the potential field. As illustrated in Fig. \ref{TPMvsOurs}(j), applying discrete differentiation in the $x$-directions reveals patterns that closely resemble the conductivity distribution in form, as shown in Fig. \ref{TPMvsOurs}(d). This gradient perspective is critical, as it highlights sharp variations in the potential field near object boundaries, enabling the potential changes to be correlated with the underlying conductivity contrasts. 

Motivated by this observation, we incorporate a physics-based processioning strategy, wherein a classical reconstruction method is first applied to generate a coarse conductivity estimate $\Sigma^p$, which is then fed into the supervised forward network. This not only provides an initial approximation of the spatial conductivity pattern but also enhances the interpretability and generalization capacity of the subsequent learning process. The formulation is given by:
\begin{equation}
\Sigma^p=\left( \mathbf{J}^T\mathbf{W}\mathbf{J}+\eta ^2\mathbf{R} \right) ^{-1}\mathbf{J}^T\mathbf{W}\Delta \mathrm{V}=\mathbf{M}_r\cdot \Delta \mathrm{V}
\end{equation}
here, $\mathbf{J} \in \mathbb{R}^{M \times N}$ is the Jacobian (sensitivity) matrix relating conductivity perturbations to boundary voltage measurements.$\mathbf{J}^\top$ denotes its transpose. $\mathbf{W} \in \mathbb{R}^{M \times M}$ is a diagonal weighting matrix. $\mathbf{R} \in \mathbb{R}^{N \times N}$ is the regularization matrix, constructed using the NOSER prior\cite{cheney1990noser}. $\eta > 0$ is the regularization hyperparameter controlling the trade-off between data fidelity and prior smoothness. $\mathbf{M}_r \in \mathbb{R}^{N \times M}$ is the reconstruction matrix that maps boundary voltage data to a coarse conductivity estimate.

\subsubsection{Gradient perspective and network design}
Predicting the discrete potential distribution is inherently challenging due to its far more complex spatial variations compared to conductivity distributions. These variations arise from the interplay of boundary conditions and multiple conductivity regions, making potential distributions visually difficult to interpret or segment. To address this challenge, we experimented with numerous backbone networks \cite{NIPS2012_c399862d,he2016deep,woo2018cbam}, but their performance was suboptimal. This is because most of these networks were originally designed for image classification tasks, focusing on extracting semantically representative features, but falling short in capturing both local details and global contrast information essential for potential distribution prediction.

Ultimately, we drew inspiration from U²-Net\cite{qin2020u2}, a nested U-structured network designed for salient object detection. U²-Net excels at extracting multi-scale contextual information while maintaining resolution, making it particularly well-suited for predicting potential distributions. Our forward network adopts a similar nested architecture but incorporates modifications to tailor it to the unique characteristics of potential fields. Specifically, all convolutional layers in our network utilize 3$\times$3 kernels, and all max-pooling layers employ 2$\times$2 kernels, ensuring consistent spatial scaling and efficient feature extraction. Additionally, we introduce physics-informed guidance to enhance feature extraction while making the network lightweight and efficient\cite{chen2024enhancing}. This modified architecture, shown in Fig.\ref{F-NET}, effectively balances detail preservation and global context extraction, addressing the unique challenges of potential field prediction.

\begin{figure*}[htbp]
\centering
\includegraphics[scale=1]{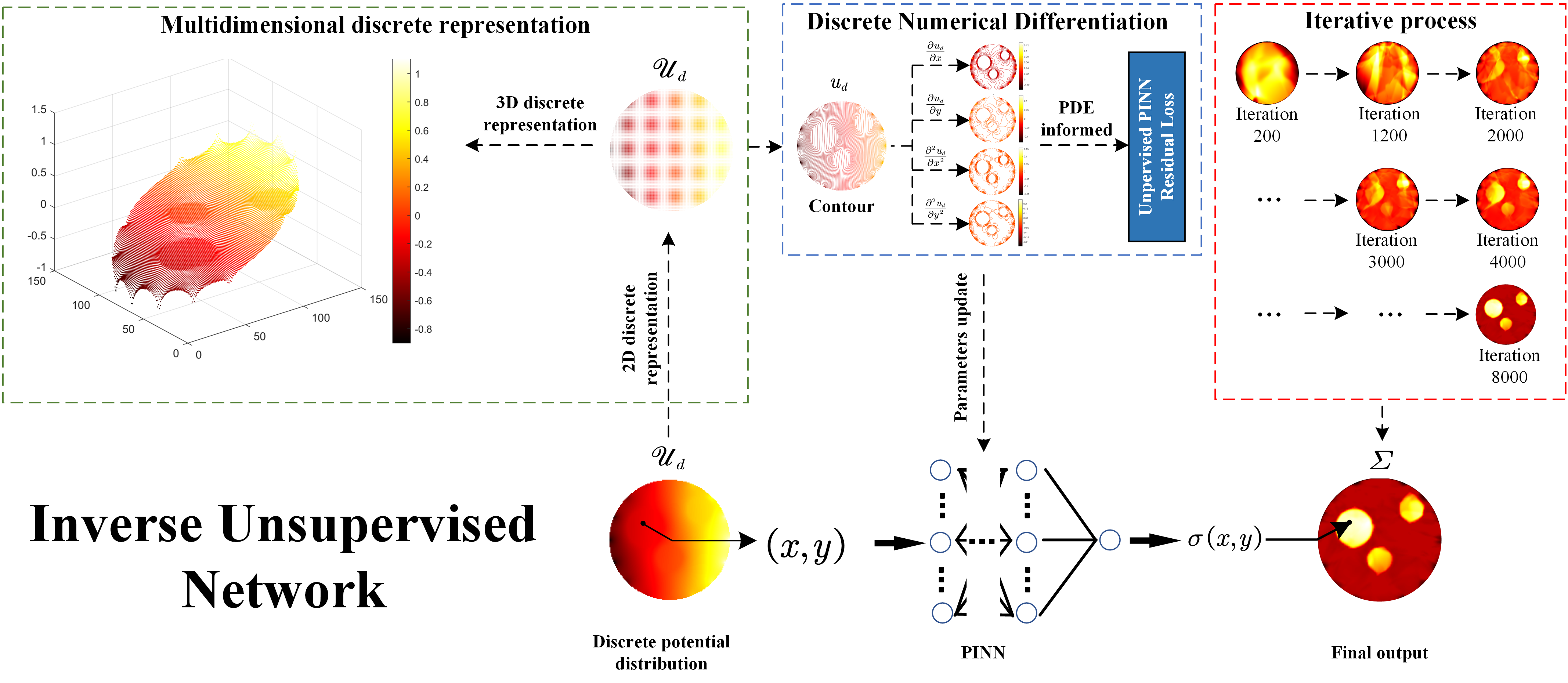}
\caption{The architecture of
inverse unsupervised network.}
\label{I-Net}
\end{figure*}
\subsection{Inverse unsupervised networks}
In this section, we introduce the model-driven inverse unsupervised network $\mathscr{F}_\sigma$, designed to reconstruct the conductivity distribution $\sigma(x, y) $ within the ROI $\mathscr{H} \subset \mathbb{R}^2$, as shown in Fig. \ref{I-Net}. The network takes the coordinates of each point in $\mathscr{U}_d$ as input and uses the discrete numerical gradients of $\mathscr{U}_d$ to enforce PDE constraints, ensuring that the solution $\sigma(x, y) $ converges to a physically valid reconstruction. In addition, the network is inspired by the work of Wang et al. \cite{wang2021understanding}, which is a multi-layer perceptron composed of four hidden layers, each of which has 64 neurons and tanh activation function, and residual connections are added to each layer to deepen the network and enhance its performance. Below, we provide a detailed explanation of the loss terms that constitute the inverse PINN network.

\subsubsection{PDE constraint loss}
The PDE constraint ensures that the predicted conductivity satisfies the governing physics described by Equation (1). On the discrete $128\times128$ grid, the PDE residual is evaluated numerically as
\begin{equation}
\mathscr{L}_{PDE}^d = \nabla \cdot \left( \sigma_d \nabla u_d \right),
\end{equation}
where the gradients of $u_d$ are approximated by central finite differences, e.g.,
$\partial u_d/\partial x \approx (u_d(x+h,y)-u_d(x-h,y))/(2h)$, where $h=1$ and likewise in the $y$-direction. These are multiplied element-wise with $\sigma_d$ to form $\sigma_d\nabla u_d$, and the divergence is computed by again applying central differences. This formulation avoids analytic differentiation and remains valid under piecewise continuous conductivity fields, allowing the framework to handle sharp boundaries and non-smooth distributions.

\subsubsection{Boundary Condition Loss}
To enforce the Neumann boundary conditions from Equation (2), we introduce the following term: 
\begin{align}
\mathscr{L}_{Neumann}^b = \sigma_b \frac{\partial u_b}{\partial n_b}
\end{align}
where this term ensures that the predicted conductivity at the boundary satisfies the relationship between the normal derivative of the potential field and the injected current.

Additionally, we incorporate a boundary conductivity loss to further enhance the accuracy and consistency of conductivity predictions at the domain boundaries:
\begin{align}
\mathscr{L}_\sigma^b = \sigma_b - \sigma_{\partial \mathscr{H}_b}^{*}
\end{align}
where $\sigma_{\partial \mathscr{H}_b}^*$ is the ground truth or expected conductivity values at the boundary. This term constrains the predicted conductivity to align with prior knowledge or experimental data, ensuring that the boundary conditions are consistent with physical expectations. 

\subsubsection{Regularization loss}
To improve the smoothness, sparsity, and physical realism of the reconstructed conductivity distribution, we employ an isotropic Total Variation (TV) regularization term \cite{gonzalez2017isotropic}. On the discrete grid, this is computed as 
\begin{align}
\mathscr{L}_{TV} = \sqrt{(\nabla_x \sigma_d)^2 + (\nabla_y \sigma_d)^2 + \xi},
\end{align}
where $\nabla_x \sigma_d$ and $\nabla_y \sigma_d$ are central finite-difference approximations of the conductivity gradients, and a small parameter $\xi>0$ is added for numerical stability. Compared with anisotropic TV, which penalizes horizontal and vertical variations separately, the isotropic form ensures rotation invariance and yields smoother reconstructions in the presence of noise or sharp conductivity interfaces.

Additionally, a non-negativity constraint (Hinge Loss) is incorporated to ensure that the predicted conductivity remains physically realistic by enforcing non-negativity:
\begin{align}
\mathscr{L}_{hinge} = \max(0, 1 - \sigma_h)
\end{align}
Finally, a parameter regularization term is introduced to penalize large model weights, thereby preventing overfitting and improving the generalization ability of the network: 
\begin{align}
\mathscr{L}_{parameter} = \|w_\sigma\|^2
\end{align}
which penalizes large model weights, helping to prevent overfitting and improve generalization.
\subsubsection{Overall inverse loss}
The final inverse loss function integrates all the individual loss components, weighted by their respective coefficients:
\begin{align}
\mathscr{L} _{inv}&=\frac{\alpha}{\mathscr{H}}\sum_{d\in \{\mathscr{H} \}}{\left( \mathscr{L} _{PDE}^{d} \right) ^2}+\frac{\beta}{T}\sum_{t\in top_T\mathscr{L} _{\mathrm{PDE}}}{\left| \mathscr{L} _{PDE}^{d} \right|}\nonumber\\
                  &+\frac{\gamma}{\left| \partial \mathscr{H} _b \right|}\sum_{b\in \partial \mathscr{H} _b}{\left| \mathscr{L} _{Neumann}^{b} \right|}+\frac{1}{|\partial \mathscr{H} |}\sum_{b\in \partial \mathscr{H} _b}{\left| \mathscr{L} _{\sigma}^{b} \right|}\nonumber\nonumber\\
                  &+\frac{\mu}{|\mathscr{H} \cup \partial \mathscr{H} |}\sum_{h\in \{\mathscr{H} \cup \partial \mathscr{H} \}}{\mathscr{L} _{hinge}}\nonumber\\
                  &+\frac{\lambda}{|\mathscr{H}|}\sum_{d\in \mathscr{H}}{\mathscr{L} _{TV}}\nonumber\\
                  &+\varrho \mathscr{L} _{parameter}
\end{align}
Here, the weighting coefficients $\alpha, \beta, \gamma, \lambda, \mu,$ and $\varrho$ are hyper-parameters that balance the influence of each loss term. The second term in Equation (15) corresponds to a \textit{selective PDE residual minimization}, where only the top‑T grid points with the largest residuals are penalized ,which focuses optimization on regions with the most significant physical inconsistencies, thereby improving convergence stability and reconstruction accuracy.

It is worth noting that several components of this composite loss—such as the PDE residual term and the Neumann boundary constraint—are inspired by the prior works of  \cite{bar2021strong} and \cite{pokkunuru2023improved} , who introduced similar strategies for enforcing physical consistency in EIT. However, our formulation is not a direct copy: instead, we selectively combine and adapt the most effective elements from both works, while discarding components less suited for practical settings. This selective integration is reflected in our hyperparameter design and tuning, as detailed below.  

\subsubsection{Hyper-parameters tuning}
The full-inverse EIT problem is highly nonlinear and ill-posed, meaning that even slight changes in parameters can result in dramatic variations in the solution. After extensive testing and optimization, we finalized the hyperparameter settings for our proposed approach framework.To highlight our methodological differences, Table \ref{hyperparameters} compares our choices with those of Bar et al. and Pokkunuru et al. While some common structures are retained (e.g., PDE residual and TV regularization), our framework departs in several important ways:  
(i) electrode-related hyperparameters $\delta$ and $\epsilon$, which cannot be directly measured in practice, are discarded in our formulation;  
(ii) the isotropic TV term is emphasized with a larger weight $\lambda$, improving robustness against noise and sharp boundaries;  
(iii) the Neumann constraint is tuned with $\gamma=1.5$ rather than a fixed value of 1.

\begin{table*}[!htbp]
    \caption{Comparison of hyper-parameters for different methods.}
    \centering
    \setlength{\tabcolsep}{2.5pt}
    \begin{tabular}{@{}c*{10}{c}@{}}
        \toprule
        \multirow{2}{*}{\textbf{Method}} & \multicolumn{10}{c}{\textbf{Hyper-parameters}} \\ 
        \cmidrule(l){2-11}
         & $\alpha$ & $\beta$ & $T$ & $\gamma$ & $\delta$ & $\epsilon$ & $\varrho$ & $\lambda$ & $\mu$ & $\xi$\\ 
        \midrule
        Bar et al.\cite{bar2021strong}   & 0.01 & 0.01 & 40   & 1    & 1    & 1    & 1e-8 & 0.01 & 0    & 0    \\
        Pokkunuru et al.\cite{pokkunuru2023improved}  & 0.05 & 0.05 & 40   & 1    & 0.1  & 100  & 1e-6 & 0.01 & 10   & 0\\
        Ours          & 0.01 & 0.01 & 40   & 1.5    & 0  & 0  & 1e-6 & 1 & 8   & 1e-4\\
        \bottomrule    
    \end{tabular}
    \label{hyperparameters}
\end{table*}

\subsection{Pseudo code}
The pseudo-code of the proposed approach is shown in Algorithm 1.

\section{Numerical and experimental studies}
\begin{figure*}[htbp]
\centering
\includegraphics[scale=0.8]{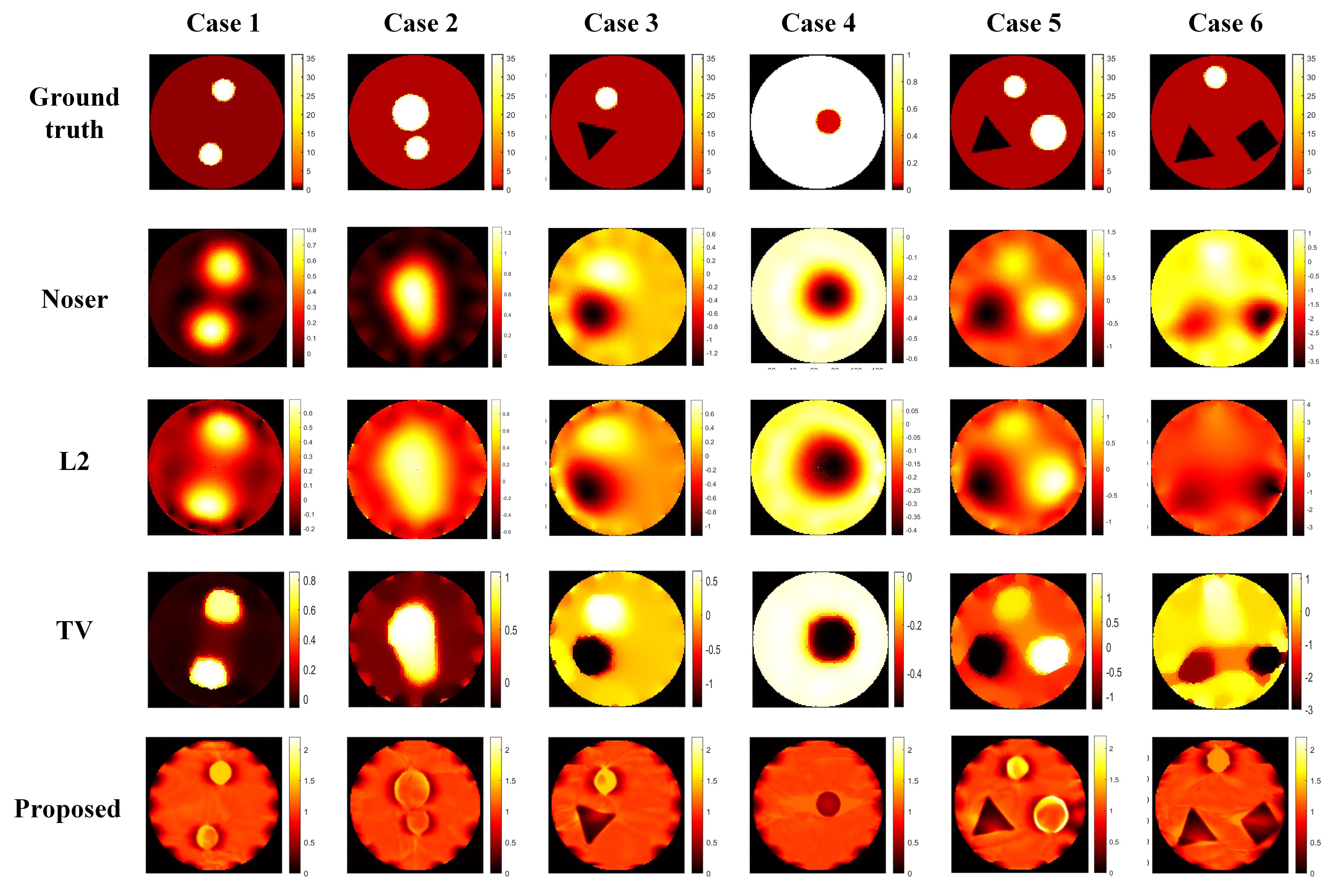}
\caption{Full-inverse EIT reconstructions with simulated data.}
\label{Simulation}
\end{figure*}

\begin{figure*}[htbp]
\centering
\includegraphics[scale=0.85]{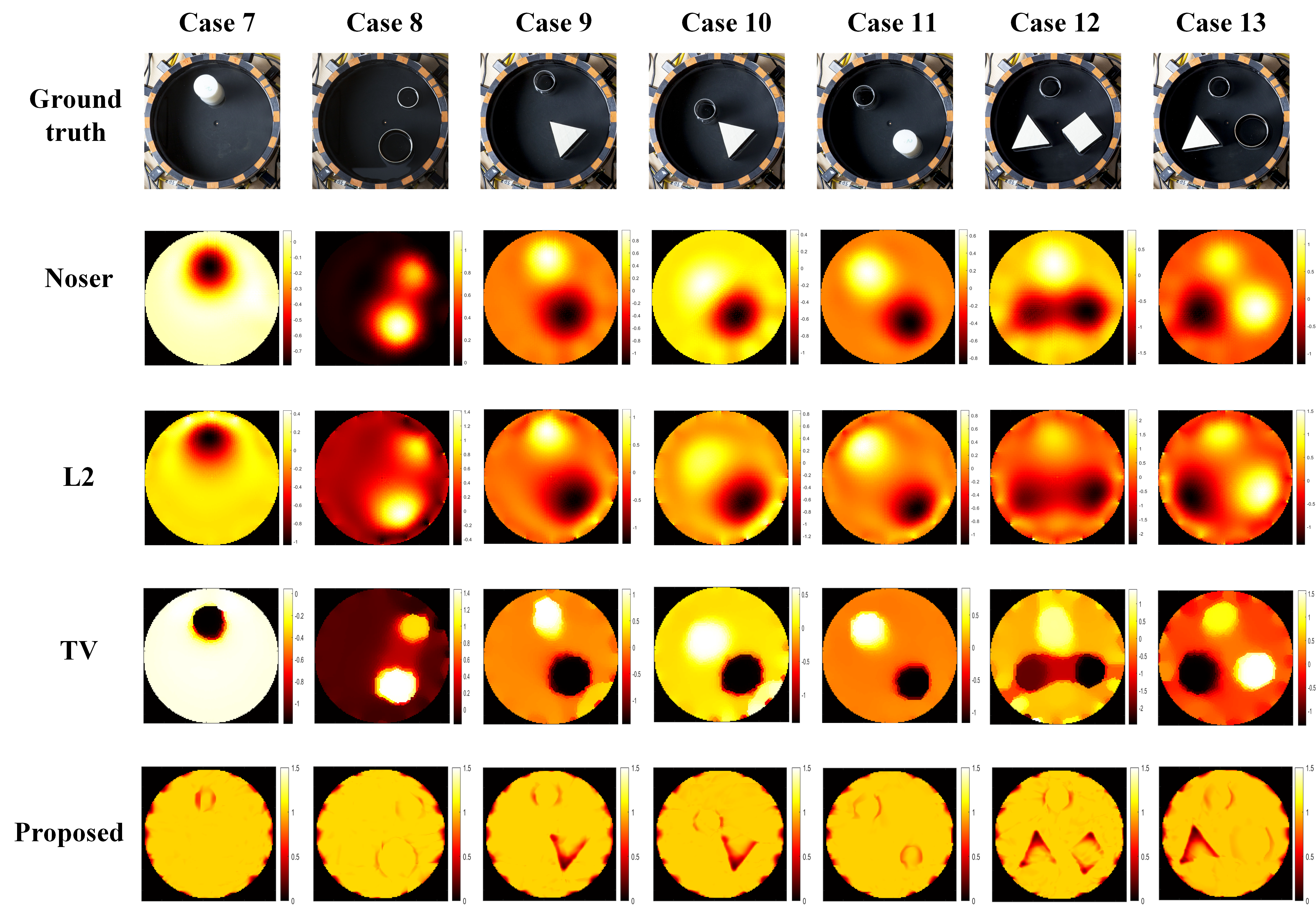}
\caption{Full-inverse EIT reconstructions with experimental water tank.}
\label{Experiment}
\end{figure*}

This section presents the simulation and real-world experiments conducted to validate the effectiveness of proposed approach for full-inverse EIT. The proposed approach framework was implemented in Python, with the TensorFlow library used to develop both the forward and inverse networks. All experiments were performed on a desktop equipped with two NVIDIA GTX 3090 GPUs. To evaluate the performance of the proposed approach, we conducted both simulation and experimental tests. Specifically, six simulation cases (cases 1–6, as shown in Fig. \ref{Simulation}) were designed, and for the experimental tests, a series of plastic and metal objects with various shapes were placed in a cylindrical tank with a diameter of 28 cm filled with saline water to create diverse conductivity distributions, as illustrated in Fig. \ref{Experiment}. The electrodes, current stimulation patterns, and measurement modes in the experimental tests were consistent with those used in the simulations. Experimental data (cases 7–13)\cite{hauptmann2017open} were collected using the KIT-4 measurement system\cite{kourunen2008suitability}, further demonstrating the applicability of the proposed approach in realistic scenarios.

\subsection{Dataset and training}
We utilized EIDORS\cite{adler2006uses} to generate the dataset for our experiments. A unit circle with a radius of 1 was modeled, incorporating a 16-electrode adjacent measurement pattern. The internal discrete potential distributions were generated under the boundary conditions defined in Equation (2). The finite element mesh consisted of 7744 triangular elements, corresponding to a pixel grid size of 128×128. The dataset included shapes such as circles, equilateral triangles, and squares, with their positions randomly generated. The dataset categories comprised the following configurations: a single circle, two circles, one equilateral triangle plus one circle, one equilateral triangle plus two circles, and one equilateral triangle combined with one circle and one square. Each category contained 3000 samples. The radii of the circles were randomly generated between 0.1 and 0.28, the side lengths of the equilateral triangles ranged from 0.5 to 0.65, and the squares had side lengths between 0.4 and 0.5. To simulate measurement noise, Gaussian noise with a signal-to-noise ratio (SNR) randomly varying between 40 dB and 60 dB was added to the voltage measurements.

To evaluate the proposed approach under extreme conditions, we increased the contrast between the background conductivity and the conductivity of the objects, simulating scenarios such as the presence of metallic objects. This significantly amplified the non-smoothness of the conductivity distribution, with conductivities randomly generated between 0.05 and 36 (excluding the background conductivity of 1). The dataset was split into 80\% training data, 10\% validation data, and 10\% test data.

\subsection{Comparison methods and evaluation indices}
To provide a more intuitive demonstration of the effectiveness of the proposed approach, we compared our method with three traditional approaches that incorporate different types of prior information: the NOSER prior\cite{cheney1990noser}, the Tikhonov prior ($L2$ norm prior)\cite{vauhkonen1998tikhonov}, and the Total Variation (TV) prior\cite{gonzalez2017isotropic}. As discussed in the \textbf{Introduction}, the works of L. Bar et al. and A. Pokkunuru et al. were unable to achieve full-inverse EIT under non-smooth conductivity distributions. Consequently, it was not feasible to test their methods in this context for a direct comparison. We also quantitatively evaluated the performance of each method using three metrics: the Structural Similarity Index (SSIM), the Correlation Coefficient (CC), and the Relative Error Index (RIE). A higher SSIM value indicates greater similarity between the reconstructed and true conductivity distributions. The CC measures the linear relationship between the reconstructed and actual conductivity distributions, with a higher CC indicating a closer match. The RIE quantifies the relative error in the reconstruction, where a higher RIE value indicates a larger discrepancy between the reconstructed and true distributions.

\subsection{Simulation results}
The reconstruction results using simulation data are shown in Fig. \ref{Simulation}. The first row represents the true conductivity distributions, while the second, third, and fourth rows correspond to the reconstructions using the NOSER, $L2$, and TV algorithms, respectively. The final row displays the results of our proposed approach. From the comparison of the reconstruction results, it is evident that traditional methods introduce significant artifacts, whereas the proposed approach achieves a clear reconstruction of object boundaries and sharp corners, demonstrating its superior performance. Notably, the proposed approach also achieves the best evaluation metrics among all methods, as shown in Table \ref{simu-eval}. It is worth highlighting that our approach performs well even under extreme conditions, where the conductivity contrast between the objects and the background is very large (background conductivity = 1, object conductivity = 36). This underscores the proposed approach’s ability to handle highly non-smooth conductivity distributions effectively. However, it should be noted that the inverse network in the proposed approach is computationally intensive, requiring 7000 to 10,000 iterations for convergence. The primary focus of this work is to address the limitations of L. Bar et al. and A. Pokkunuru et al. in reconstructing non-smooth conductivity distributions and achieving full-inverse EIT. Accelerating the inverse network's convergence remains an open challenge and will be the focus of our future research.

\subsection{Experimental results}
This section presents the application of the proposed approach to achieve full-inverse EIT using real-world physical data, a scenario not previously addressed in TPM. Fig. \ref{Experiment} illustrates the reconstruction results from real-world experiments, where the inclusions consisted of both metal and plastic objects. Similar to the simulation results, traditional methods failed to distinguish the specific shapes of the internal objects, whereas the proposed approach successfully reconstructed the shapes and contours of the inclusions with greater clarity. Moreover, the proposed approach demonstrated the ability to differentiate between metallic and plastic inclusions. For example, darker colors in the reconstruction represent plastic, while lighter colors indicate metal. It is worth noting that, due to the noise present in real voltage data, the reconstruction quality of the proposed approach in real-world experiments is slightly inferior to that in simulations. However, this is expected. Despite the noise, the proposed approach significantly outperformed traditional algorithms in accurately reconstructing the shapes and positions of inclusions, as evident in Table \ref{exper-eval}.

\begin{table*}[htbp]
\centering
\caption{Evaluation metrics for Cases 1 to 6.}
\begin{tabular}{cccccccccccccc}
\toprule
\multirow{2}{*}{\textbf{Algorithm}}
 & \multicolumn{3}{c}{Case 1} & \multicolumn{3}{c}{Case 2} & \multicolumn{3}{c}{Case 3} & \multicolumn{3}{c}{Case 4}\\
\cmidrule(lr){2-4} \cmidrule(lr){5-7} \cmidrule(lr){8-10}\cmidrule(lr){11-13}
 & SSIM & CC & RIE & SSIM & CC & RIE & SSIM & CC & RIE & SSIM & CC & RIE \\
\midrule
NOSER     & 0.3369 & 0.6785 & 1.0176 & 0.3919 & 0.7558 & 0.5188 & 0.1897 & 0.3010 & 3.6218 & 0.6246 & 0.8271 & 0.2210\\
$L_2$     & 0.2487 & 0.5512 & 1.7021 & 0.2284 & 0.5308 & 1.3868 & 0.1966 & 0.2779 & 3.1372 & 0.5299 & 0.7624 & 0.3415\\
TV        & 0.3822 & 0.6752 & 0.9859 & 0.2782 & 0.6713 & 1.0751 & 0.2144 & 0.2985 & 3.6693 & 0.6295 & 0.7709 & 0.2830\\
Proposed      & 0.4545 & 0.7523 & 0.9675 & 0.4326 & 0.7569 & 0.6935 & 0.5280 & 0.5157 & 2.5022 & 0.7417 & 0.9262 & 0.1455\\
\bottomrule
\end{tabular}

\begin{tabular}{ccccccccccc}
\toprule
\multirow{2}{*}{\textbf{Algorithm}}
 & \multicolumn{3}{c}{Case 5} & \multicolumn{3}{c}{Case 6}  \\
\cmidrule(lr){2-4} \cmidrule(lr){5-7} 
 & SSIM & CC & RIE & SSIM & CC & RIE  \\
\midrule
NOSER     & 0.2269 & 0.5172 & 1.5083 & 0.1782 & 0.2596 & 4.1265  \\
$L_2$     & 0.2248 & 0.5064 & 1.5190 & 0.2167 & 0.2560 & 2.2984 \\
TV        & 0.2721 & 0.5845 & 1.3841 & 0.2012 & 0.2614 & 3.5919  \\
Proposed      & 0.3551 & 0.7811 & 1.0853 & 0.4193 & 0.5818 & 2.2428 \\
\bottomrule
\end{tabular}
\label{simu-eval}
\end{table*}

\begin{table*}[htbp]
\centering
\caption{Evaluation metrics for Cases 7 to 13.}
\begin{tabular}{cccccccccccccc}
\toprule
\multirow{2}{*}{\textbf{Algorithm}}
 & \multicolumn{3}{c}{Case 7} & \multicolumn{3}{c}{Case 8} & \multicolumn{3}{c}{Case 9} & \multicolumn{3}{c}{Case 10}\\
\cmidrule(lr){2-4} \cmidrule(lr){5-7} \cmidrule(lr){8-10}\cmidrule(lr){11-13}
 & SSIM & CC & RIE & SSIM & CC & RIE & SSIM & CC & RIE & SSIM & CC & RIE \\
\midrule
NOSER     & 0.6354 & 0.8351 & 0.2354 & 0.4026 & 0.7654 & 0.5423 & 0.1786 & 0.3159 & 3.3578 & 0.1656 & 0.3650 & 3.2156 \\
$L_2$     & 0.5139 & 0.7520 & 0.3546 & 0.2365 & 0.5297 & 1.2693 & 0.1923 & 0.2861 & 3.1355 & 0.1846 & 0.2756 & 3.0362 \\
TV        & 0.6455 & 0.7895 & 0.2642 & 0.2956 & 0.6659 & 1.1123 & 0.2065 & 0.2856 & 3.6494 & 0.1955 & 0.2698 & 3.6038 \\
Proposed      & 0.6312 & 0.8962 & 0.2123 & 0.4287 & 0.7499 & 0.7890 & 0.3956 & 0.4896 & 2.8634 & 0.3731 & 0.4550 & 2.9920 \\
\bottomrule
\end{tabular}

\begin{tabular}{ccccccccccc}
\toprule
\multirow{2}{*}{\textbf{Algorithm}}
 & \multicolumn{3}{c}{Case 11} & \multicolumn{3}{c}{Case 12} & \multicolumn{3}{c}{Case 13} \\
\cmidrule(lr){2-4} \cmidrule(lr){5-7} \cmidrule(lr){8-10}
 & SSIM & CC & RIE & SSIM & CC & RIE & SSIM & CC & RIE \\
\midrule
NOSER     & 0.3213 & 0.6697 & 1.0320 & 0.1651 & 0.2687 & 4.0248 & 0.2156 & 0.5023 & 1.5125 \\
$L_2$     & 0.2236 & 0.5601 & 1.7009 & 0.2034 & 0.2565 & 2.2652 & 0.2338 & 0.5154 & 1.5003 \\
TV        & 0.3771 & 0.6460 & 0.9623 & 0.1988 & 0.2813 & 3.1365 & 0.2598 & 0.5607 & 1.3512 \\
Proposed      & 0.3811 & 0.7167 & 0.9635 & 0.3985 & 0.5125 & 2.2603 & 0.3645 & 0.7031 & 0.9658 \\
\bottomrule
\end{tabular}
\label{exper-eval}
\end{table*}

\section{Discussion}
\subsection{Compare with baselines}

\begin{figure}[htbp]
\centering
\includegraphics[scale=1.4]{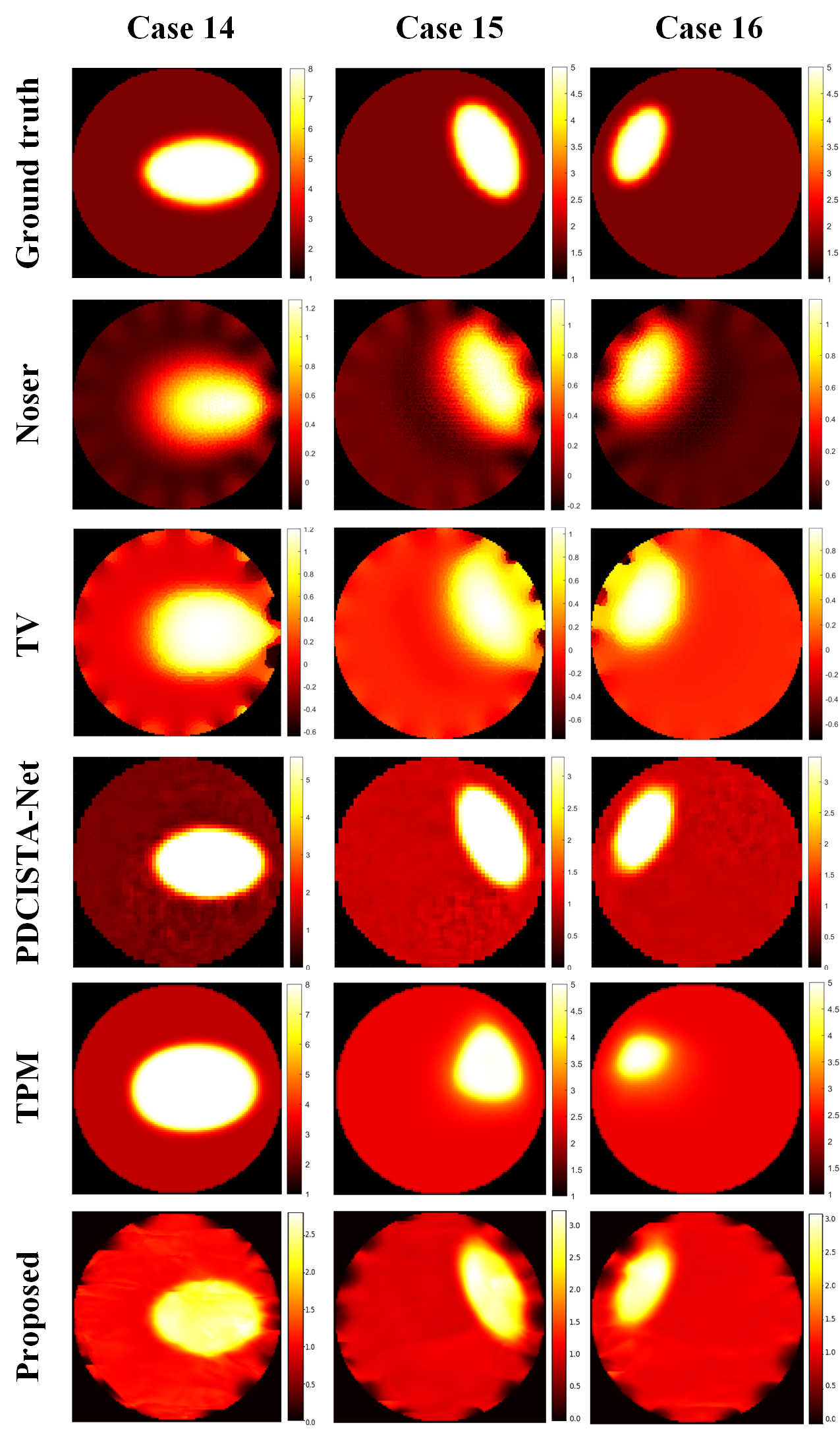}
\caption{The proposed approach reconstruction compared with baselines. }
\label{baseline}
\end{figure}

In the context of full-inverse EIT, the reconstruction results of A. Pokkunuru et al.’s TPM method\cite{pokkunuru2023improved} are highly unstable, with most cases failing to distinguish the imaging targets. For a fair comparison, we selected the only three successful cases provided in their work for EIT image reconstruction, as shown in Fig.\ref{baseline}.
The results indicate that our proposed method produces significantly fewer artifacts than traditional approaches such as NOSER and TV, while providing more accurate shape reconstruction than TPM. This validates the effectiveness of our approach in addressing the challenges of full-inverse EIT. To further evaluate performance against state-of-the-art purely data-driven methods, we additionally compared our approach with PDCISTA-Net\cite{ma2024pdcista}, one of the most advanced EIT reconstruction networks. PDCISTA-Net was trained on the same dataset and under identical experimental settings as our method. As shown in Table \ref{baselinetable}, our approach achieves superior visual reconstruction quality and higher quantitative metrics (SSIM, CC, RIE), demonstrating both better generalization and robustness under realistic experimental conditions.

\begin{table*}[h!]
\caption{Evaluation metrics for Cases 14 to 16.}
\begin{tabular}{ccccccccccc}
\toprule
\multirow{2}{*}{\textbf{Algorithm}}
 & \multicolumn{3}{c}{Case 14} & \multicolumn{3}{c}{Case 15} & \multicolumn{3}{c}{Case 16} \\
\cmidrule(lr){2-4} \cmidrule(lr){5-7} \cmidrule(lr){8-10}
 & SSIM & CC & RIE & SSIM & CC & RIE & SSIM & CC & RIE \\
\midrule
NOSER     & 0.1821 & 0.9281 & 0.8554 & 0.1754 & 0.8786 & 0.8248 & 0.1731 & 0.8513 & 0.8400 \\
TV     & 0.2013 & 0.8976 & 0.8511 & 0.1870 & 0.8415 & 0.8248 & 0.1812 & 0.8231 & 0.8392 \\
PDCISTA-Net\cite{ma2024pdcista}        & 0.6002 & 0.9086 & 0.4268 & 0.7724 & 0.9985 & 0.3364 & 0.7491 & 0.9189 & 0.3489 \\
TPM\cite{pokkunuru2023improved}      & 0.3789 & 0.9152 & 0.2824 & 0.4425 & 0.9124 & 0.3237 & 0.4412 & 0.8684 & 0.3643 \\
Proposed      & 0.6065 & 0.9219 & 0.6468 & 0.7901 & 0.9381 & 0.3175 & 0.7535 & 0.9289 & 0.3451 \\

\bottomrule
\end{tabular}
\label{baselinetable}
\end{table*}

\subsection{Influence of Boundary Excitation Conditions}
\begin{figure}[htbp]
\centering
\includegraphics[scale=0.8]{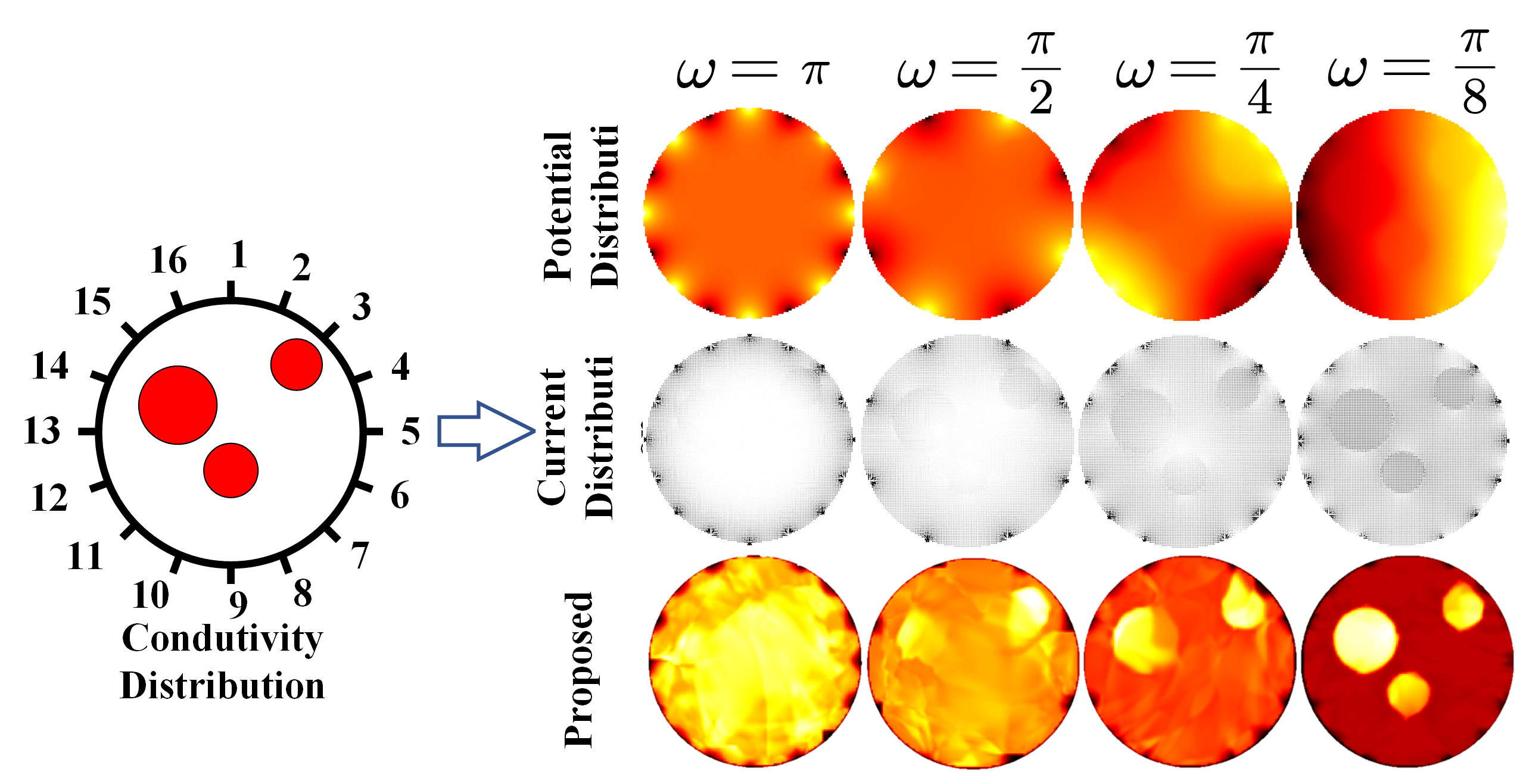}
\caption{The internal potential distribution, current distribution, and the proposed approach reconstruction results under different Neumann boundary conditions. }
\label{frequency}
\end{figure}

To analyze the effect of boundary conditions, we considered a more realistic 16-electrode setup rather than the idealized continuous current injection used in previous PINN-based studies. Specifically, we investigated the influence of excitation frequency on current penetration and image reconstruction. As shown in Fig.\ref{frequency}, increasing the excitation frequency reduces the penetration depth of the injected current, thereby limiting the amount of information available from the internal potential field. Since the inverse network in our framework is constrained by the PDE relationship between conductivity and potential, insufficient penetration directly degrades the reconstruction performance. This phenomenon is clearly reflected in the reconstructions, where higher frequencies result in notable quality deterioration. Based on these observations, we selected $\omega=8$ as the optimal frequency, balancing penetration depth and image fidelity. 

\subsection{Noise Tolerance Analysis}
\begin{figure}[htbp]
\centering
\includegraphics[scale=1]{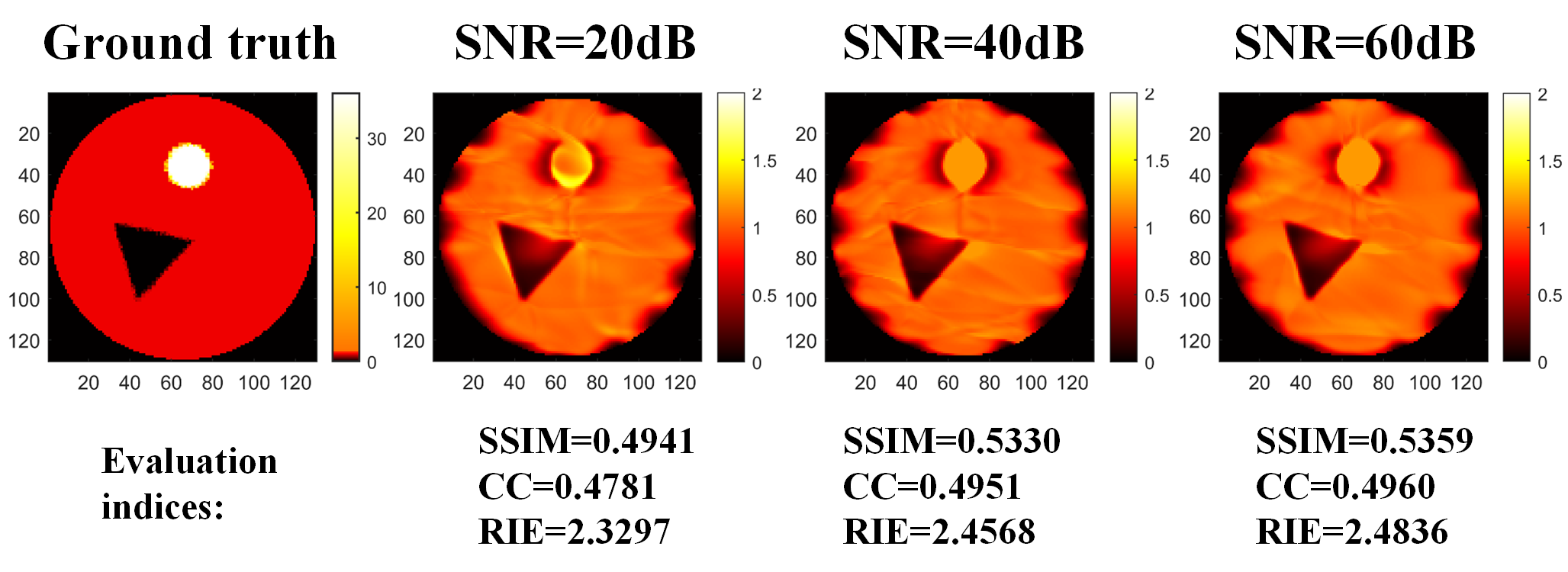}
\caption{the proposed approach reconstruction results under different noise levels. }
\label{noise}
\end{figure}

To visually demonstrate the noise resilience of the proposed approach, we randomly selected a voltage measurement dataset and added Gaussian noise to it, achieving PSNR values of 20dB, 40dB, and 60dB. The resulting noisy data was used for image reconstruction using the proposed approach, and evaluation metrics were computed, as shown in Fig. \ref{noise}. As observed, the reconstructed images remained almost unaffected as the PSNR decreased, indicating the excellent noise robustness of the proposed approach. Moreover, our real-world experiments also serve as a robustness analysis, as the KIT4 data acquisition system used in the experiments inherently introduces noise into the measurements.

\subsection{Conductivity Impact on Reconstructions}

\begin{figure}[htbp]
\centering
\includegraphics[scale=1]{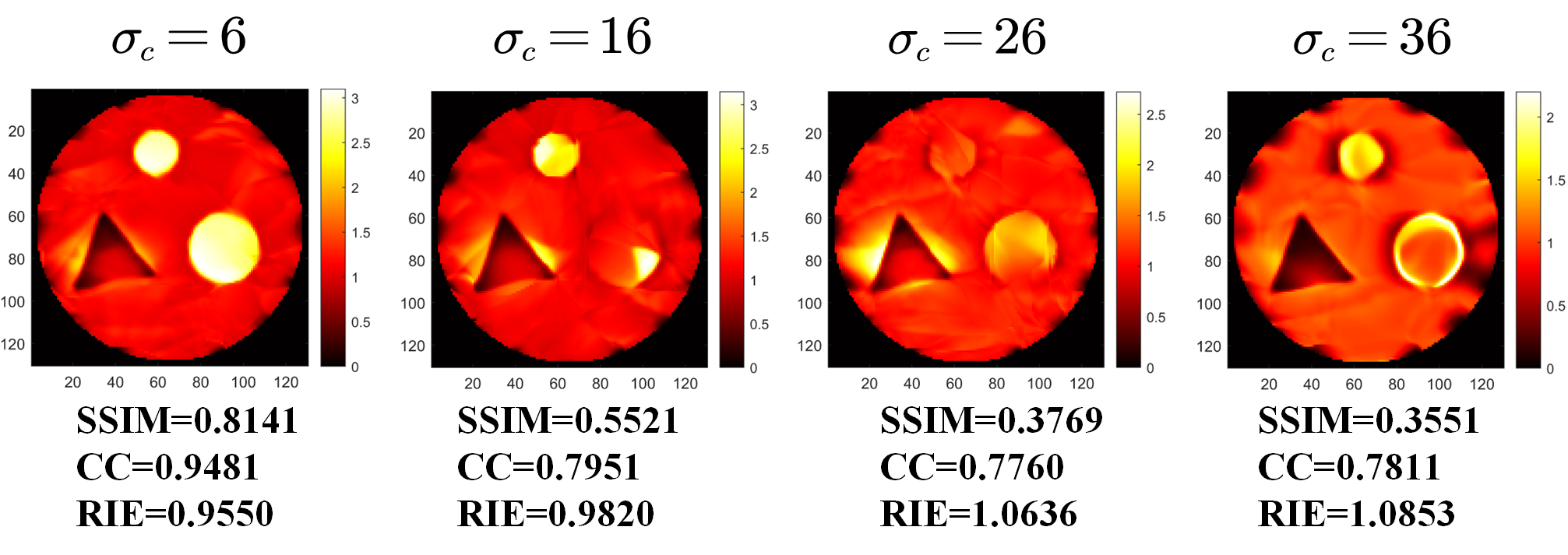}
\caption{The proposed approach reconstruction results under different conductivity. }
\label{diffduct}
\end{figure}

To evaluate the reconstruction performance of our proposed method under extreme conductivity contrasts, we conducted a parametric study using the Case 5 configuration. The background conductivity was maintained at 1, while the target circular inclusion's conductivity was systematically varied across four orders of magnitude ($\sigma=6,16,26,36$). As demonstrated in Fig.\ref{diffduct}, the reconstruction exhibits optimal accuracy at $\sigma=6$, with progressively degraded image fidelity as conductivity contrast increases. This degradation stems from intensified nonlinear perturbations in the potential field when the target-to-background conductivity ratio exceeds critical thresholds, introducing pronounced nonlinearities into the inverse problem. For quantitative benchmarking, Table \ref{tab:comparison_diff_cond} compares the maximum conductivity ratios successfully reconstructed by our method against state-of-the-art approaches, including Bar et al. and Pokkunuru et al.
\begin{table*}[h!]
    \caption{Benchmark comparison of maximum reconstructible conductivity ratios.}
    \centering
    \begin{tabular}{@{}c*{4}{c}@{}}
        \toprule
        \multirow{2}{*}{\textbf{Work}} & \multicolumn{4}{c}{\textbf{Content}} \\ 
        \cmidrule(l){2-5}
         & Background Conductivity & Maximum Conductivity & Effective Imaging & Maximum number of Phantoms \\ 
        \midrule
        \cite{bar2021strong}   & 1 & 5  & Yes & 2 \\
        \cite{pokkunuru2023improved}  & 1 & 15 & No  & 3 \\
        Ours          & 1 & 36 & Yes & 3 \\
        \bottomrule    
    \end{tabular}
    \label{tab:comparison_diff_cond}
\end{table*}

\begin{table*}[h!]
    \caption{Comparison of different methods.}
    \centering
    \begin{tabular}{@{}c*{5}{c}@{}}
        \toprule
        \multirow{2}{*}{\textbf{Work}} & \multicolumn{5}{c}{\textbf{Content}} \\ 
        \cmidrule(l){2-6}
         & Semi-inverse EIT & Full-inverse EIT & Simulation & Experiment & Non-smooth conductivity distributions\\ 
        \midrule
        \cite{bar2021strong}   & Yes & Yes(They claimed) & Yes   & No    & No   \\
        \cite{pokkunuru2023improved}  & Yes & No & Yes   & No    & No\\
        Ours          & Yes & Yes & Yes   & Yes    & Yes \\
        \bottomrule    
    \end{tabular}
    \label{tab:hyperparameters}
\end{table*}

\subsection{Study limitations and future work}

Furthermore, unlike traditional PINN-based EIT, which requires idealized continuous boundary current sources and smooth conductivity assumptions, our approach introduces a CNN-based forward predictor to handle realistic point-electrode excitations and non-smooth conductivity distributions. The CNN component serves as a data-driven prior that enables robust potential estimation under sparse and noisy boundary measurements, while the subsequent PINN stage enforces governing physics and eliminates the pure black-box nature. This hybrid mechanism reduces the reliance on over-simplified mathematical assumptions and paves the way toward future practical hardware-ready EIT imaging systems.

\begin{algorithm}[!htbp]
\caption{The proposed approach framework}
\begin{algorithmic}[1]
\Require Training dataset $\{ (\Delta \mathrm{V}^i, \mathscr{U}_d^i) \}_{i=1}^N$, boundary conditions $\zeta$, loss coefficients $\alpha, \beta, \gamma, \lambda, \mu, \varrho$, grid step size $h$
\Ensure Reconstructed conductivity distribution $\varSigma = \{ \sigma(x, y) \mid (x, y) \in \mathscr{H} \}$
\State  \textbf{Stage 1: Forward Supervised Network}
\State  Define forward network $\mathscr{F}_u$ parameterized by $\theta = \{ W, b \}$
\State  Train $\mathscr{F}_u$ to minimize the supervised loss:
\[
\hat{\theta} = \mathrm{arg}\min_{\theta} \frac{1}{N} \sum_{i=1}^N \left\| \mathscr{F}_u(\Delta \mathrm{V}^i) - \mathscr{U}_d^i \right\|_2^2
\]
\State  Output discrete potential field $\mathscr{U}_d = \mathscr{F}_u(\Delta \mathrm{V})$

\State  \textbf{Stage 2: Numerical Differentiation on $\mathscr{U}_d$}
\State  Compute discrete gradients: $\frac{\partial u_d}{\partial x}, \frac{\partial u_d}{\partial y}$ 
\State  \textbf{Stage 3: Inverse Unsupervised Network}
\State  Define inverse network $\mathscr{F}_\sigma$ parameterized by $\phi = \{ w_\sigma \}$
\State  Initialize $\phi$ randomly
\State  Define inverse loss $\mathscr{L}_{inv}$
\State  Train $\mathscr{F}_\sigma$ to minimize $\mathscr{L}_{inv}$
\While{not converged}
    \State  Update $\phi$ using gradient descent:
    \[
    \phi \leftarrow \phi - \eta \nabla_{\phi} \mathscr{L}_{inv}
    \]
\EndWhile
\State  Output reconstructed conductivity distribution:
\[
\varSigma = \{ \sigma(x, y) \mid (x, y) \in \mathscr{H} \}
\]
\end{algorithmic}
\end{algorithm}

\section{Conclusion}
In this paper, we proposed a physics-embedded dual-learning imaging
framework for EIT combining supervised network and unsupervised network to address the challenges of full-inverse electrical impedance tomography (EIT). Our method decouples the forward and inverse problems, overcoming the limitations of traditional PINN frameworks that rely on smooth conductivity assumptions and struggle with full-inverse EIT under non-smooth conditions. By replacing automatic differentiation with discrete numerical differentiation, the proposed approach effectively handles real-world scenarios involving sharp boundaries and highly non-smooth conductivity distributions. The forward CNN network in the proposed approach accurately predicts discrete potential distributions, capturing intricate spatial variations even under extreme conductivity contrasts, while the inverse PINN network ensures physically plausible conductivity reconstructions by adhering to the underlying PDE constraints. The method demonstrated strong robustness to noise, maintaining high-quality reconstructions even at low signal-to-noise ratios, and outperformed traditional algorithms, such as NOSER, Tikhonov, and TV priors, in both simulated and real-world experiments. The proposed approach was successfully applied to full-inverse EIT in real-world scenarios for the first time, clearly reconstructing the shapes and contours of inclusions and distinguishing between metallic and plastic materials. Despite the computational intensity of the inverse network, which requires multiple iterations for convergence, the proposed approach's promising performance opens the door for further optimization, and its robustness and scalability make it suitable for a wide range of practical applications. The proposed dual-learning design not only relaxes the smoothness requirement of conductivity and eliminates idealized boundary assumptions, but also significantly improves interpretability compared with traditional end-to-end CNN inversion. This framework provides a practical pathway toward clinically and industrially deployable PINN-enabled EIT.

\bibliographystyle{cas-model2-names}


\bibliography{ref.bib}



\end{document}